\documentclass[12pt]{iopart}

\usepackage{epsfig}
\usepackage{array}
\begin{document}

\letter{}
\title{Strangeness Production at RHIC in the Perturbative Regime}
\author{Daphne Y. Chang\dag,
 Steffen A.~Bass\dag\ddag,
and
 Dinesh K. Srivastava\S,
}
\address{\dag\ Department of Physics, Duke University, 
             Durham, North Carolina 27708-0305, USA}
\address{\ddag\ RIKEN BNL Research Center, Brookhaven National Laboratory, 
             Upton, New York 11973, USA}            
\address{\S\ Variable Energy Cyclotron 
Centre, 1/AF Bidhan Nagar, Kolkata 700 064, India}  

\ead{dyc3@phy.duke.edu}

\submitto{\JPG}
\pacs{25.75.-q,12.38.Mh}

\begin{abstract}
We investigate strange quark production in Au-Au collisions at RHIC in
the framework of the Parton Cascade Model(PCM).  The yields of (anti-) strange
quarks for three production scenarios -- primary-primary scattering, full 
scattering, and full production -- are compared to a proton-proton 
baseline.  Enhancement of strange quark yields 
in central Au-Au collisions compared to scaled
p-p collisions increases with the number of secondary interactions.
The centrality dependence of strangeness production for the three 
production scenarios is studied as well.  For all production mechanisms,
the strangeness yield increases with $(N_{\rm part})^{4/3}$.  The perturbative
QCD regime described by the PCM is able to account for
%
up to 50\% of the 
observed strangeness at RHIC.  
\end{abstract}

\maketitle

The investigation of strangeness production in relativistic heavy ion collisions
has been proven to be a powerful tool for the study of highly excited nuclear matter, both
in terms of the reaction
dynamics and in terms of its hadrochemistry
\cite{Rafelski:pu,koch86,koch88,Braun-Munzinger:1994xr,Letessier:ad,Vance:1999pr,Soff:1999et,Braun-Munzinger:1999qy,Rafelski:2001hp,Braun-Munzinger:2001ip}.
Furthermore, strangeness has  been suggested as a signature
for the creation of a Quark-Gluon-Plasma (QGP) \cite{Rafelski:pu,koch86,koch88}: 
in the final state of a heavy-ion collision, subsequent to the formation 
and decay of a QGP, strangeness has been predicted to
be enhanced relative to the strangeness yield in
elementary hadron-hadron collisions. 
%
%
Strangeness production is suppressed
in purely hadronic collisions due to 
the large constituent strange quark mass, whereas (anti-)strange
quarks are nearly massless in a QGP due to the restoration of chiral symmetry and are therefore
produced in abundance.
The chemical equilibration times for strange and multistrange particles have been shown to
be considerably shorter
in the QGP phase than in a thermally equilibrated hadronic fireball.
The dominant strangeness production mechanism, i.e. $gg \to s\overline{s}$, should allow
for equilibration times similar to the expected QGP lifetime.  

Recent data from the Relativistic Heavy-Ion Collider (RHIC) at Brookhaven Lab have provided strong
evidence for the existence of a transient QGP -- among the most exciting findings are
strong (hydrodynamic) collective flow \cite{HuKoHe,teaney,PHENIXv2,Esumi,STAR-v2,STAR-v2b}, 
the suppression of high-$p_T$ particles \cite{Wa1,GyViWa,PHENIX-su,Adams:2003kv} 
and evidence for parton recombination as hadronization mechanism at intermediate 
transverse momenta \cite{FrMuNoBa1,GrKoLe1,Vo,Nonaka:2003hx}. 

However, the dynamics of initial strangeness production in the deconfined phase at RHIC, how
it relates to the data, and what we can learn from it, has not
received  much attention lately. In this letter, we address this topic and
investigate strangeness production in the framework of a microscopic transport model --
the Parton Cascade Model -- which is well suited for the investigation of the
pQCD driven early deconfined reaction phase of a ultra-relativistic heavy-ion collision.

\section*{The Parton Cascade Model}

The parton cascade model (PCM) provides a detailed
space-time description of nuclear collisions at high energy,
from the onset of hard interactions among the partons of the
colliding nuclei up to the moment of hadronization~\cite{GM}.
The PCM is best suited for the description of the early, pre-equilibrium
reaction phase, which is dominated by perturbative hard scattering
processes. Due to the introduction of a momentum cut-off, needed to regularize
the IR-divergence of the perturbative parton-parton cross sections, the PCM
is not 
%
%
equipped to account for soft
 (re)scatterings,
 which would dominate
a thermalized ensemble of quarks and gluons. 

The PCM is based on the relativistic Boltzmann equation for the time
evolution of the parton density due to perturbative QCD interactions:
\begin{equation}
p^\mu \frac{\partial}{\partial x^\mu} F_i(x,p) = {\cal C}_i[F] \, .
\end{equation}
The collision term ${\cal C}_i$ is a nonlinear functional of the
phase-space distribution function $F(x,p)$, containing the matrix
elements which account for the following processes:
\begin{equation}
\label{processes}
\begin{array}{lll}
g g \to g g \quad&\quad (g g \to q \bar q)
  &\quad q g \to q g \\
q q' \to q q' \quad&\quad q q \to q q
  &\quad (q \bar q \to q' \bar q') \\
q \bar q \to q \bar q
  &\quad q \bar q \to g g \quad& \\
q g \to q \gamma \quad&\quad q \bar q \to \gamma \gamma
  & \quad q \bar q \to g \gamma
\end{array}
\end{equation}
with $q$ and $q'$ indicating different quark flavors.  
The processes in parentheses can be optionally disabled to exclude
production due to them (see later).
The
corresponding scattering cross sections are expressed in terms
of spin- and color-averaged amplitudes $|{\cal M}|^2$~\cite{Cutler.78}:
\begin{equation}
\label{dsigmadt}
\left( \frac{{\rm d}\hat \sigma}
     {{\rm d} Q^2}\right)_{ab\to cd} \,=\, \frac{1}{16 \pi \hat s^2}
        \,\langle |{\cal M}|^2 \rangle
\end{equation}
The total cross section, necessary for the transport calculations, is
obtained from
(\ref{dsigmadt}):
\begin{equation}
\label{sigmatot}
\hat \sigma_{ab}(\hat s) \,=\,
\sum\limits_{c,d} \, \int\limits_{(p_T^{\rm min})^2}^{\hat s}
        \left( \frac{{\rm d}\hat \sigma }{{\rm d} Q^2}
        \right)_{ab\to cd} {\rm d} Q^2 \quad .
\end{equation}
The low momentum-transfer cut-off $p_T^{\rm min }$ regularizes
the infrared divergence of the parton-parton cross section.  
For the initial
parton distribution, we choose the GRV-HO parameterization \cite{grv} and sample the
distribution functions at the initialization scale $Q_0^2$. For our calculation we
choose $Q_0^2=(p_T^{\rm min})^2=$~0.5~GeV$^2$.   Initial
state partons are virtual, i.e. their momentum distribution is 
governed by the parton distribution function, yet they propagate
with the velocity of their parent hadron.
                                                                                
Additionally, we include the branchings $q \to q g$, 
$ g \to gg$ and $g \to q\overline{q}$ \cite{frag}.  The soft and
collinear singularities in the showers are avoided by terminating the
branchings when
the virtuality of the time-like partons drops below $\mu_0 = 1$ GeV.
The present work is based on the thoroughly revised, corrected, and
extensively tested implementation of the original parton cascade 
model \cite{vni},
called VNI/BMS~\cite{VNIBMS}.

The PCM distinguishes between two types of strangeness -- intrinsic 
strangeness and produced strangeness.  Intrinsic strangeness arises
from strange quarks that are already contained 
in the Dirac sea of the initial
parton distribution, and are released when they are put
on-shell through scattering.  These scattering processes can either be
elastic (anti-)quark -- (anti-)quark (e.g. $qs \to qs$) or 
elastic (anti-)quark -- gluon (i.e. $gs \to gs$) interactions.
Produced strangeness refers to $s \bar s$ pairs newly created in
binary collisions, such as
$q\overline{q} \to s\overline{s}$ and 
$gg \to s\overline{s}$ as well as in time-like branching processes,
e.g.  $g \to s\overline{s}$.
Strange quarks produced from these processes constitute the 
{\em enhancement} of strangeness in the system.

\section*{Dynamics of Strangeness Production}

Our analysis is set up to investigate the following scenarios:
\begin{itemize}
\item {\bf primary-primary scattering:} all partons are allowed to scatter
only once. Interactions are limited to elastic scattering, no annihilation or
pair production is permitted (i.e. the processes in parentheses in 
table \ref{processes} are disabled), 
and time-like branching processes are also disabled. 
 This calculation will allow for the analysis of the 
{\em release} of strangeness from the initial state through
``first-chance'' energetic initial state parton scattering. 
\item {\bf full scattering:} partons are allowed to rescatter multiple
times. Interactions are again limited to elastic scattering without
time-like branching (i.e. the processes in parenthesis in 
table \ref{processes} 
%
%
remain disabled). 
Analysis of this calculation will provide an estimate for
the amount of strangeness released from the initial state in secondary 
interactions.
\item {\bf full production:} partons may rescatter both elastically as
well as inelastically, and undergo time-like branchings in the final state.
This calculation allows for both the initial state release as well as 
the additional production of $s \bar s$ pairs, and therefore provides an
estimate of the total strangeness production in the pQCD regime accessible
by the PCM.
\end{itemize}
None of the above described calculations contain hadronization. 
We evaluate the $s \bar s$ distributions at the level of quarks  --- these can
 then
be compared to the $s$ and $\bar s$ valence quark content of measured 
hadron distributions.

Figure~1 shows the rapidity distribution for all three scenarios
as well as for the $s \bar s$ distribution of the initial state.
Central Au+Au collisions at $\sqrt{s}_{NN}=$~200~GeV are shown with
full symbols, whereas the open symbols refer to a sample of 
minimum bias p-p collisions at the 
same energy. The p-p 
%
%
yield has been scaled by 197: if the Au-Au
calculations were just an independent superposition of 197 nucleon + 197 nucleon
collisions, the two curves would fall on top of each other.

\begin{figure} 
\centerline{\epsfig{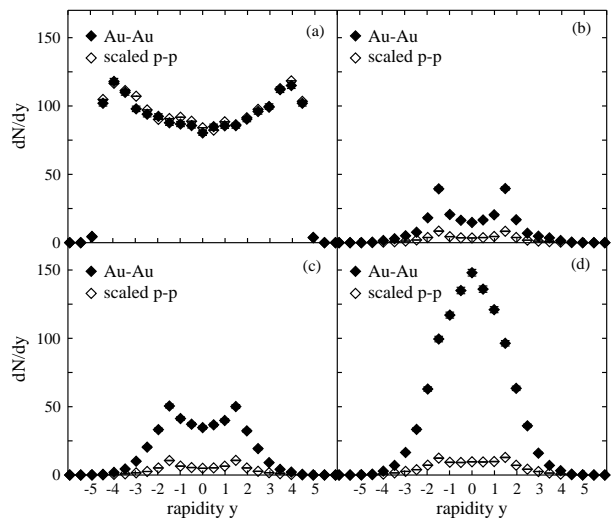}} 
\caption{Distribution of (anti-) strange quarks as a function of rapidity y
for central Au-Au(full symbols) and p-p (open symbols) collisions at 
$\sqrt{s}_{NN}=$~200~GeV.  Proton-proton results are scaled by
$N_{\rm part,Au-Au}/N_{\rm part,p-p}=197$.(a) shows the initial state strange
(sea) quark
distribution.  (b) shows a calculation with primary-primary scattering,
whereas all scattering processes (primary-primary, primary-secondary, and 
secondary-secondary) are turned on in (c).  (d) shows the full production
calculation  where in addition to the (re-)scattering processes (c) strangeness
production is enabled via inelastic binary collisions and time-like branching.}
\label{fig1}
\end{figure} 

Frame (a) shows the rapidity distribution of the initial $s$ and $\bar s$ 
(sea) quarks contained in the parton distributions of the 
Au-Au and p-p systems, which
serves as the baseline for our strangeness enhancement analysis. 
Since the partons in the initial state are virtual, their rapidities
are ill-defined. However, 
%
%
 one can relate
a parton's longitudinal momentum $p_z$ before or
immediately after the first collision of a parton  to 
the rapidity variable $y = Y + \ln x + \ln(M/Q_s)$,
where $Y$ is the rapidity of the fast moving nucleon, $M$ is the
nucleon mass, and $Q_s$ denotes the typical transverse momentum
scale. Depending on the picture of the initial state, $Q_s$ is
either given by the average intrinsic virtuality, often called the
saturation scale \cite{sat-Q}, or by the typical transverse momentum
given to the parton in the first interaction which brings it onto
the mass shell. In any case, $|\ln(M/Q_s)| < 1/2$ for Au-Au
collisions at RHIC.
This initial state
baseline can now be compared to
frame (b), showing the primary-primary scattering calculation, frame (c), 
showing  the full scattering calculation and  frame (d), which
displays the full production calculation.

The release of 
intrinsic strangeness via scattering can therefore be studied by comparing
frames (b) and (c), and the 
enhancement of strangeness via $s \bar s$ production can be seen by
comparing frame (d) with frame (c). 
We find that the rapidity distribution of released strangeness
%
%
%
%
still has the double-hump structure of the initial state, where the 
positions of the peaks are shifted by several units of rapidity.
 Approximately 13\% of the initial strangeness is 
released by primary-primary interactions, with another 10\% being set
free through primary-secondary and secondary-secondary rescatterings.
Newly created  $s \bar s$ pairs, however, are mostly produced around 
mid-rapidity and exhibit a roughly Gaussian rapidity distribution.
For Au-Au collisions, a comparison between frames (c) and (d) 
shows that  about 40\% of the total strangeness yield is 
released from the initial state, while nearly 60\% is newly produced.  
The total strangeness yield ($s$ + $\bar s$) of the Au-Au PCM calculation 
accounts for 55\% of STAR's measured yield at
mid-rapidity \cite{STAR}. This undersaturation of strangeness production
in the PCM is most likely due to the limitation of the model to pQCD
processes. Previous calculations based on pQCD rate-equations \cite{biro93a} or
a very early implementation of the Parton Cascade Model \cite{geiger93a} confirm
this trend.

The enhancement of strangeness production in Au-Au reactions compared to 
scaled p-p collisions increases from primary-primary scattering (b)
to full scattering (c) to full production (d). The enhancement factors 
for the integrated yield in the three scenarios are, respectively, 
4.5, 6.0, and 10.5.  We attribute this increase in the strangeness yield 
enhancement to be primarily caused by the 
enhanced probability of parton rescattering
in Au-Au versus p-p.  
%
%

\begin{figure} 
\centerline{\epsfig{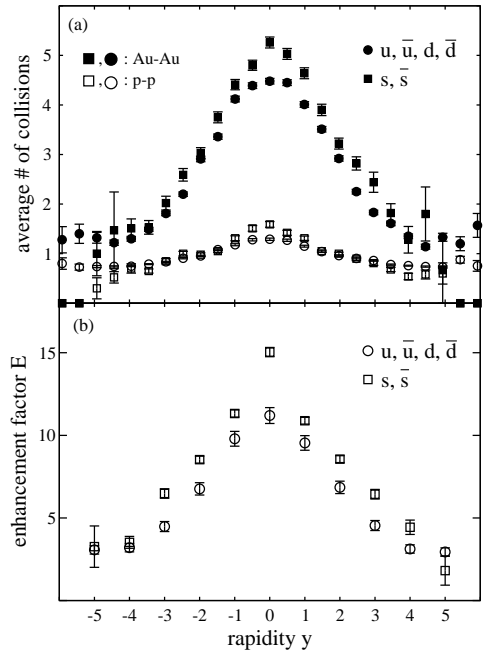}} 
\caption{(a) Average number of collisions a parton undergoes as a function of 
rapidity y for central Au-Au (full symbols) and p-p (open symbols) collisions at 200
GeV.  Strange and anti-strange quarks (squares) collide more often than (anti-) up and down
quarks (circles) in Au-Au collisions, with the enhancement in collision number 
greatest at mid-rapidity. (b) Rapidity dependence of enhancement
factors $E_{\rm u, \overline{u}, d, \overline{d}}$(circles), 
and $E_{\rm s, \overline{s}}$(squares), defined
as the relative enhancement of the (anti-particle) yields in central Au-Au collisions compared to
the scaled yields in p-p collisions in the respective rapidity bin.}
\label{fig2}
\end{figure}

A common way of comparing strangeness yields and enhancement across
different systems is by calculating the Wroblewski factor \cite{wrob}, 
defined as:
\begin{equation}
\lambda_{s} \equiv \frac{2 \langle s \bar s\rangle}{\langle u \bar u\rangle + \langle d \bar d \rangle}
\end{equation}
In our analysis we only use the multiplicities of those partons
which have been released or produced through a binary scattering 
or a branching process. The Wroblewski factors extracted for the
different modes of our calculations can be found in table~1. 
For the full Au-Au calculation we find a Wroblewski factor of 
0.5. However, this result has to be interpreted very cautiously and
cannot be directly compared to the experimental findings, since
our extraction of the Wroblewski factor neglects all hadronization
effects, e.g. the splitting of  gluon into quark-anti-quark pairs
at hadronization. Furthermore, it remains unclear whether all 
released or produced quarks in the PCM will be converted into 
valence quarks of hadrons at hadronization.

\begin{table}
\setlength{\extrarowheight}{6pt}
\centerline{\begin{tabular}{|m{1.5in} m{.8in} m{.5in}|}
\hline & Au-Au & p-p\\
primary-primary scattering & 0.26 & 0.25\\
full scattering & 0.31 & 0.28\\
full production & 0.52 & 0.36\\
\hline
\end{tabular}}
\label{table1}
\caption{Wroblewski factor $\lambda_s$, calculated on the basis
of released/produced (anti-)strange quarks in the PCM.}
\end{table}

We quantify the amount of parton rescattering and production
enhancement in figure~2: the upper frame shows the average number of 
collisions quarks undergo as a function of rapidity
for Au-Au as well as for p-p, whereas the lower frame displays the
quark production enhancement factor, namely the ratio of the $u,d$ and $s$
quark rapidity distributions of Au-Au over the scaled p-p distributions. 
As can be seen, strangeness enhancement is largest at mid-rapidity, 
which correlates to (anti-)strange quarks that have scattered 
the most number of times. We also find
that (anti-) up and down quarks scatter less often than (anti-) strange 
quarks in 
Au-Au collisions, and that $u \bar u$ and $d \bar d$ production has a smaller 
dependence on rescattering than $s \bar s$ production.
%
%
%

The lower frame of figure~3 shows the number distribution of
binary parton-parton center of mass collision energies  
(i.e. $\sqrt{\widehat{s}}$) for the process 
$g g \to s \bar s$ and $ q \bar q \to s \bar s$
for central Au-Au (circles) and p-p (squares) 
collisions at 200 GeV.  p-p results are scaled with the ratio of 
the number of elastic $ g g \to g g$ and $ q \bar q \to g g$ 
reactions in Au-Au relative to p-p,
thus allowing us to determine in which regime
 (anti-)strange quark pair-production 
in Au-Au is enhanced (or suppressed) compared to p-p.
The scaling with respect to elastic $ g g \to g g$ reactions 
quantifies this enhancement beyond the trivial increase due to
the larger parton density in Au-Au vs. p-p.
A significant enhancement of pair production  is observed for 
for small $\sqrt{\widehat{s}}$ up to 2.5~GeV followed by a suppression 
with increasing $\sqrt{\widehat{s}}$ beyond 4~GeV.
This behavior can be explained by analyzing the $\sqrt{\widehat{s}}$
dependence of the respective elastic and inelastic $gg$ scattering
cross sections, which are shown in the upper frame of figure~3:
the cross section for (anti-) strange quark production
decreases strongly with increasing $\sqrt{\widehat{s}}$, 
while the elastic gluon-gluon cross section remains constant.  Since
the inelastic $g g \to s \bar s$ process dominates strangeness production,
a decrease in $g g \to s \bar s$ cross section with increasing 
$\sqrt{\widehat{s}}$ 
suppresses strangeness production for parton-parton collisions at 
high $\sqrt{\widehat{s}}$.  
%
%
At the same time, partons in Au-Au collisions scatter and fragment
much more frequently than in p-p collisions, and therefore
lose energy relatively quickly.  This ensures that
partonic collisions at smaller $\sqrt{\widehat{s}}$ are more likely
in nucleus-nucleus interactions.
The implication of this finding is broad: strangeness production
is driven  by the amount of {\em quasi-thermal} rescattering at small 
parton-parton $\sqrt{\widehat{s}}$ rather than by a large 
amount of available energy.

	\begin{figure}	 
	\centerline{\epsfig{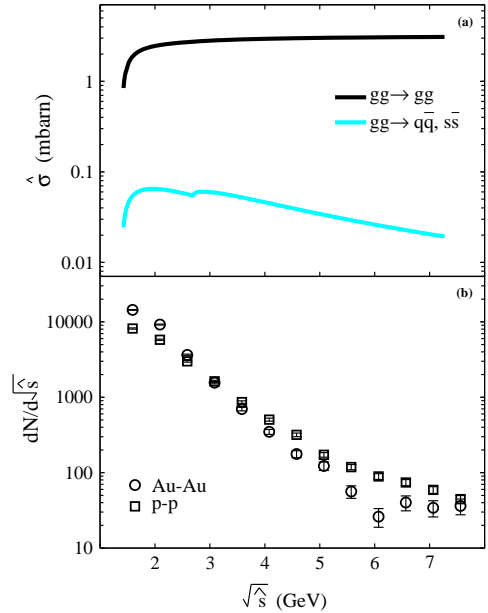}} 
	\caption{(a) Center of mass energy dependence of the g-g elastic and inelastic
cross sections.  The inelastic cross section may lead to the pair production of
(anti-) strange quarks.  The cross section for (anti-) strange quark production 
decreases with $\sqrt{\widehat{s}}$, while the elastic gluon cross section remains constant.
(b) Binary parton-parton center of mass collision energy distribution 
for the reaction $g g \to s \bar s$ and
$ q \bar q \to s \bar s$ as a 
function of $\sqrt{\widehat{s}}$ for central Au-Au (circles) and p-p (squares) 
collisions at 200 GeV.  p-p results are scaled with the ratio of 
the number of elastic $ g g \to g g$ reactions in Au-Au relative to p-p.  
A significant enhancement of pair production  is observed for 
for small $\sqrt{\widehat{s}}$ followed by a suppression 
with increasing $\sqrt{\widehat{s}}$.}
	\label{fig3}
	\end{figure}

Finally, we wish to address the centrality dependence of strangeness
production in the perturbative domain.
In our analysis, we varied the centrality of the system by two different
methods: (a) by varying the 
target/projectile size for central collisions and (b) by varying the 
impact parameter for Au-Au collisions. The results of this study
for the three production scenarios are shown in figure~4.
Full symbols represent calculations of varying target/projectile size, 
whereas stars represent calculations with varying impact parameter.  
For each of the three production scenarios, the total integrated 
strangeness multiplicity increases with $(N_{\rm part})^{4/3}$ 
(fitted curves), indicative of a scaling with binary collisions, which
should not come as a surprise given the pQCD input of our PCM calculation.

\begin{figure} 
\centerline{\epsfig{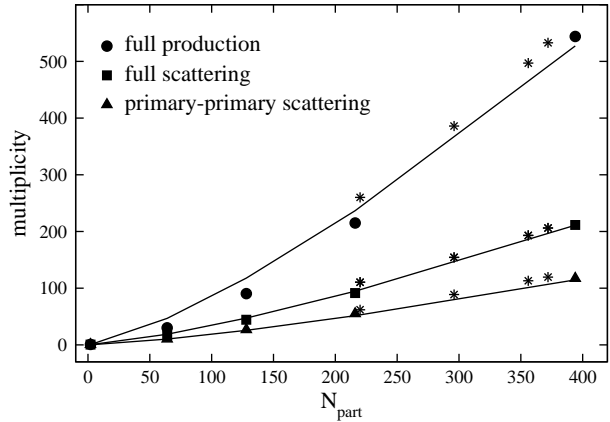}} 
\caption{Total integrated multiplicity of (anti-) strange quarks as a function of the number
of participants $N_{\rm part}$ for collision systems at $\sqrt{s}_{NN}$=~200~GeV.  Full symbols represent
central p-p ($N_{\rm part}$=2), S-S ($N_{\rm part}$=64), 
Cu-Cu ($N_{\rm part}$=128), Ag-Ag ($N_{\rm part}$=216),
and Au-Au ($N_{\rm part}$=394) collisions.  Stars are the results of Au-Au collisions at various 
impact parameters.  Full production (circles), full scattering (squares), and primary-
primary scattering (triangles) processes are the same as previously defined.  The yields of 
all three scenarios increase with $(N_{\rm part})^{4/3}$ (fitted curves).}
\label{fig4}
\end{figure}

\section*{Summary and outlook}

We have studied the production of strangeness in Au-Au collisions at RHIC in
the framework of the Parton Cascade Model(PCM).  The yields of (anti-) strange
quarks for three production scenarios --- primary-primary scattering, full 
scattering, and full production --- have been compared to a proton-proton 
baseline. In Au-Au collisions at $\sqrt{s}_{NN}$=~200~GeV, about 40\% of 
the strangeness yield
is released from the initial state, and 60\% of the yield is newly produced
through binary parton-parton interactions and final state radiation.
We find an enhancement of strange quark yields 
in central Au-Au collisions compared to scaled
p-p collisions --- this enhancement rises 
with the number of secondary interactions of the respective partons.
Strangeness production therefore seems to be sensitive to quasi-thermal rescattering.
A comparison with STAR's measurement shows that the PCM can account
for 55\% of the observed strangeness yield at mid-rapidity. 
The underprediction of the measured strangeness yield can be attributed
to the limitation of the PCM to the dynamics of the early pQCD driven
pre-equilibrium phase of the heavy-ion reaction.
This limitation is largely
due to the momentum cut-off in leading order pQCD calculation for binary cross sections
in the PCM, which prohibits soft non-perturbative scatterings.
Naturally one would expect a significant amount of strangeness to be
produced in the later thermalized QGP phase of the reaction.
A more comprehensive treatment of parton-parton interactions, e.g. via
the introduction of a screening mass instead of a hard momentum cut-off
may  allow the PCM to account for a higher fraction of the 
experimentally observed strangeness. 
In the future we shall therefore turn our attention to the investigation of
strangeness balance functions, strangeness equilibration in infinite 
partonic matter and the estimation of soft, nonperturbative contributions
to strangeness production.

\ack  
D.Y.C. and S.A.B thank Berndt M\"uller for stimulating discussions and valuable
suggestions concerning this work.
This work was supported in part by RIKEN, the Brookhaven National 
Laboratory, and DOE grants DE-FG02-96ER40945 and DE-AC02-98CH10886.
S.A.B. acknowledges support from an Outstanding Junior Investigator
Award (DOE grant DE-FG02-03ER41239).

\end{document}